\documentclass[twocolumn,showpacs,preprintnumbers,amsmath,amssymb]{revtex4}

\usepackage{graphicx}% Include figure files
\usepackage{dcolumn}% Align table columns on decimal point
\usepackage{bm}% bold math

\begin{document}

\preprint{APS/123-QED}

\title{Energy and number of collisions fluctuations  \\ in inelastic gases }

\author{R. Lambiotte}
\email{Renaud.Lambiotte@ulg.ac.be}

\author{M. Ausloos}
\email{Marcel.Ausloos@ulg.ac.be}

\affiliation{%
SUPRATECS, Universit\'e de Li\`ege, B5 Sart-Tilman, B-4000 Li\`ege, Belgium
}%
\author{J.M. Salazar}
\email{jmarcos@u-bourgogne.fr}

\affiliation{%
LRRS UMR-5613 CNRS,  Facult\'e des Sciences Mirande, 9 Av. Alain Savary 21078 Dijon Cedex , France
}%

\date{15/10/2005}

\begin{abstract}
We study by numerical simulations the two-dimensional Inelastic Maxwell Model (IMM), and show how the inelasticity of collisions together with the fluctuations of the number of collisions undergone by a particle lead to energy fluctuations that decay  like a power-law. These fluctuations are associated to a shrinking of the available phase space. We find the asymptotic scaling of these energy fluctuations and show how they  affect  the tail of the velocity distribution during long time intervals.
\end{abstract}

\pacs{89.75.Fb, 89.75.Hc, 87.23.Ge}

\maketitle

% main text
\section{Introduction}

 Inelastic gases are low density systems 
composed of macroscopic particles themselves performing {\em inelastic} collisions \cite{book,book2}. 
Due to their inelasticity, these systems dissipate kinetic energy, so that  their total energy asymptotically vanishes if they are not supplied by an external energy source. 
Nonetheless, it has been shown \cite{vanno1, brey2, ernst2, ernst3,barrat} that their velocity distribution, when homogeneous, usually reach 
a  self-similar solution, i.e. form preserving solution whose time dependence occurs through one parameter
\begin{equation}
f(v; t) = \frac{1}{\sqrt{T(t)}} f_S(\frac{v}{\sqrt{T(t)}}).
\end{equation}
$T(t)$ is the {\em granular temperature}  defined kinetically by 
\begin{equation}
   T( t) = <\frac{1}{d}\frac{m  V^{2}}{2}>
   \end{equation}
   where ${\bf V}$=${\bf v} - {\bf u} $ is the random velocity, {\bf u} is the local mean velocity and d is the dimension of the system. The average is performed with the one particle velocity distribution $f({\bf v}; t)$. 
Such scaling velocity distributions have been observed in a large variety of kinetic models, and have been shown to generically highlight  overpopulated high energy tail, whose details depend on the  model \cite{ernst2, naim1, balda1,lambi}.

In Statistical Physics, interactions usually play a mixing role whose effects is to bring the system into its equilibrium state \cite{bois}. This justifies, for instance in kinetic theory,  to use the average number of collisions in order to  measure relaxation to equilibrium. In the case of inelastic gases, on the other hand, it is well known that the dissipative collisions prevents the system to reach its equilibrium state, i.e. additionally to a mixing effect that randomises the particles velocities, inelastic collisions act as an energy sink that makes their kinetic energy decrease. 

In this paper,  we focus on the effect of this additional effect
on simplified kinetic models, the so-called Inelastic Maxwell Models (IMM) \cite{naim1,balda1,lambi, ernst,ernst0,nana}. After introducing the model in section IIA, we  perform numerical simulations, thereby showing that
the average energy of particles depends on their collision history, i.e. on the number of collisions they have undergone in the course of time. This relation that also appears in simpler one-dimensional models \cite{lambi1},  is related to the fluctuating character of the number of inelastic collisions.
We show that the resulting energy fluctuations correspond to fluctuations of the phase space volume and that they relax  like power-laws toward the asymptotic state.

\section {Collision number statistics}
\subsection{Inelastic Maxwell Model}

Usually, inelastic gases are defined as assemblies of smooth inelastic hard spheres (IHS), i.e. particles whose interactions do not transfer  angular momentum, and are instantaneous and dissipative. In the absence of external force, the grains move freely between successive collisions and undergo a collision when particles i and j are in contact, with the collision rule: 

\begin{eqnarray}
{\bf v}_{i}^{'} &=& {\bf v}_{i} - \frac{(1+ \alpha )}{2 \alpha } {\mbox{\boldmath$\epsilon$}}
({\mbox{\boldmath$\epsilon$}}.{\bf v}_{ij}) \cr
{\bf v}_{j}^{'} &=& {\bf v}_{j} + \frac{(1+ \alpha )}{2 \alpha } {\mbox{\boldmath$\epsilon$}}
({\mbox{\boldmath$\epsilon$}}.{\bf v}_{ij}) 
\end{eqnarray} 
where ${\bf v}_{ij} \equiv {\bf v}_{i} - {\bf v}_{j}$ and ${\mbox{\boldmath$\epsilon$}}$ is the unitary vector along the axis joining the centers of the two colliding
spheres, ${\mbox{\boldmath$\epsilon$}} \equiv \frac{{\bf r}_{ij}}{|{\bf r}_{ij}|}$.
The primed velocities are the velocities
before the collision, the unprimed ones are the post-collisional velocities. 
Energy dissipation is accounted through the so-called normal restitution coefficient, $\alpha$ $\in ~ ]0:1]$. $\alpha$=1 corresponds to the elastic limit. Let us also stress that we assume that the restitution coefficient $\alpha$ is constant, 
i.e. we neglect the dependance of $\alpha$ on the relative velocity of the particles
 \cite{rosa,bril}.
 
  \begin{figure}

\includegraphics[angle=-90,width=3.3in]{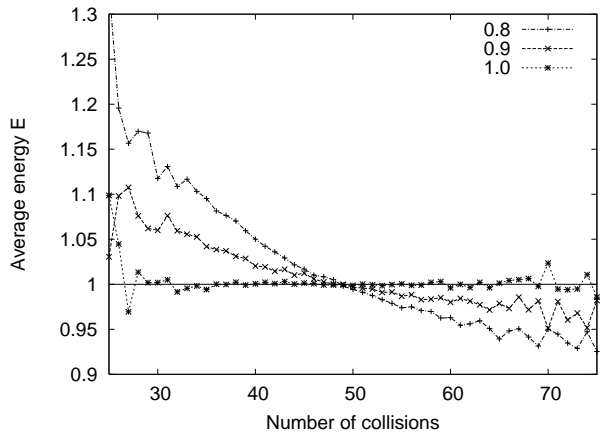}% Here is how to import EPS art

\caption{\label{figure1} Average energy of particles as a function of their number of collisions, after $50$ collision per particle. The dashed line is the constant value $<E>=1$. The system is composed of 10000000 particles with $\alpha=0.8$, $\alpha=0.9$ and $\alpha=1.0$. 
}
\end{figure}

\begin{figure}

\includegraphics[angle=-90,width=3.3in]{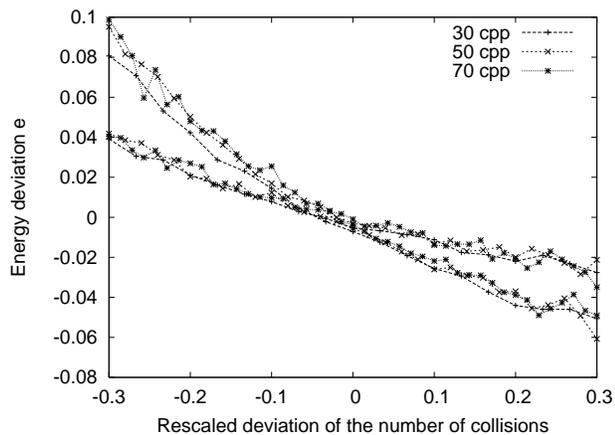}% Here is how to import EPS art

\caption{\label{figure2} Energy deviation $e$ as a function of the rescaled deviation of the number of collisions $\delta$, after 40, 70 and 100 collisions per particles (cpp).  The gas is composed of 10000000 particles with $\alpha=0.8$ and $\alpha=0.9$. 
}
\end{figure}

In the low density limit, by assuming that pre-collisional correlations may be neglected \cite{soto}, the system is  described by the inelastic Boltzmann equation. In the following, we assume that the system is and remain homogeneous.
Moreover, we use standard mean field methods in order to simplify the mathematical structure of the collision operator. To do so, we assume that the collision frequency between particles $i$ and $j$, i.e. proportional to $v_{ij}$ in the case of hard spheres, may be approximated by the mean field quantity $\sqrt{T(t)}$. This leads to the following kinetic equation:

\begin{eqnarray}
 \frac{\partial f({\bf v}_{i};t)  }{\partial t}  &=& \frac{1}{2 \pi} \int d\theta \int  d{\bf v}_{2}  ~  [\frac{1}{\alpha} 
 f({\bf v}_{i}^{'};t)  f({\bf v}_{j}^{'};t) \cr &-&  f({\bf v}_{i};t)  f({\bf v}_{j};t) ] 
 \label{kin}
\end{eqnarray} 
where $f_{i}(t) \equiv f({\bf v}_{i};t) $ and $f^{'}_i(t) \equiv f({\bf v}^{'}_1;t)$ and $\theta$ is defined by  the relation  $\cos \theta  \equiv \frac{{\mbox{\boldmath$\epsilon$}}. {\bf v}_{12}}{v_{12}}$.
Angular integrals and the factor $\sqrt{T(t)}$ have been absorbed into the time scale. This model is usually called Inelastic Maxwell Model (IMM), and several variations of it have been considered in the literature  \cite{naim1,balda1,lambi, ernst,ernst0,nana}. It is well known that it leads to an exactly solvable sets of equations for the velocity moments 
 $m_{n}(t) =<(v^2)^{n}>$,  and that its asymptotic scaling solutions is characterised by a power-law tail $v^{-\mu(\alpha)}$, where $\mu(\alpha)$ goes to infinity in the elastic limit $\alpha \rightarrow 1$. The corresponding first diverging moment of the scaling solution is found by solving a transcendental equation \cite{ernst0}. Let us also  remind the closed temperature equation $
\partial_t T = -\frac{(1-\alpha^{2})}{4} T
$
that obviously leads to an exponential decay for this quantity.

\begin{figure}

\includegraphics[angle=-90,width=3.3in]{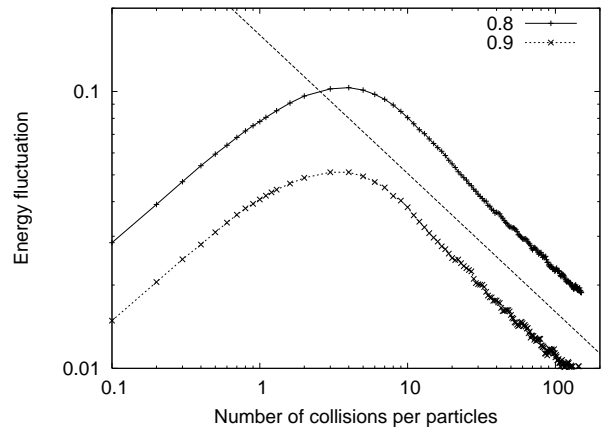}% Here is how to import EPS art

\caption{\label{figure3} Energy fluctuation $\sigma_e$ as a function 
of the number of collisions per particles.  The gas is composed of 10000000 particles with $\alpha=0.8$ and $\alpha=0.9$. 
}
\end{figure}

\subsection{Power-law relaxation and scaling}

In order to explore the energy statistics of the particles, we have performed Direct Simulation Monte-Carlo (DSMC) simulations of the IMM \cite{bird}. They are based on the stochastic  interpretation of the kinetic equation Eq.\ref{kin}
and consists in picking randomly pairs of colliding particles at each step.
 It is straightforward to show that the number of inelastic collisions $N$ undergone by a particle at a given time $t$ fluctuates. This  quantity is distributed according to the Poisson law $\frac{t^N e^{-t}}{N!}$, where, by construction, $t$ is the average number of collisions suffered by particles. Let us remind that the Poisson distribution satisfies
 
 \begin{eqnarray}
 <N>_t &=& t \cr
 \sigma \equiv \sqrt{\frac{<N^2>_t -  <N>_t^2}{<N>_t^2}} &=& t^{-\frac{1}{2}}.
 \label{poisson}
 \end{eqnarray}

In order to evaluate the effect of these fluctuations on the particles energies, we have started simulations from a Maxwell-Boltzmann initial condition, and let the simulation run during $t$ collisions per particle.
At that time, the simulation is paused, the particles velocities are rescaled so that the average energy $<E>=1$ and we measure the average energy $<E>_{\Delta}$ of particles having performed $N=t+\Delta$ collisions. The results (Fig.\ref{figure1}) clearly show that the average energy of particles is a decreasing function of their number of collisions, thereby showing that ensembles of particles, discriminated by their  number of collisions, are characterised by different quantities of energy on average. This feature comes from the fact that particles having undergone less collisions have also dissipated less energy due to inelastic collisions. Simulations also show that this relation obviously does not take place when the collisions are elastic, as expected. When $\alpha< 1$, the following scaling expression is found for long times $t>10$ and small values of $\delta \equiv \Delta/t$ (Fig.\ref{figure2})

\begin{equation}
e(\Delta) = - r(\alpha) ~ \frac{\Delta}{t} + ...
\label{scaling}
\end{equation}
where $e(\Delta)$ is the deviation to the average defined by $e \equiv <E>_{\Delta} - 1$ and $r(\alpha)$ is a positive parameter that vanishes in the limit $\alpha \rightarrow 1$. For small times, in contrast, we observe a breaking of this relation, and a non-negligeable constant at $\Delta=0$
\begin{equation}
 e(\Delta) = e(0) - e^{'}(0) \Delta  + ....
 \label{scaling2}
 \end{equation}
  Let us note that $e(0)$ increases with $t$ for small times, reaches a maximum around $t=5$ and decreases very fast later on.

Relations \ref{poisson} and  \ref{scaling}  imply that the fluctuations of $<E>_{\delta}$ decrease asymptotically like power-laws
\begin{equation}
\sigma_e= \sqrt{<e^2>} \sim - r(\alpha) ~ t^{-\frac{1}{2}}.
\label{scaling2}
\end{equation}
We observe this behaviour by DSMC (Fig.\ref{figure3}), namely $\sqrt{<e^2>}$ decreases like the power-law $t^{-\frac{1}{2}}$ after an  increase that takes place during some mean collision times. This initial increase is due to the breaking of relation \ref{scaling} for small times, and to the non-vanishing value of $e(0)$.
In the case of elastic collisions $\alpha=1$, $\sqrt{<e^2>}$ is strictly zero, i.e. the dynamics are restrained on the initial constant energy surface. In contrast, when $\alpha<1$,  the initial increase of  $<e^2>$ corresponds to a broadening of the available volume in phase space, that is followed by a power-law shrinking of this volume. 
Let us stress that these fluctuations relax on much slower time scales than  the usual exponential relaxations  taking place in kinetic theory. 

Amongst others, these fluctuations contribute to the rapid emergence of fat tails in the velocity distribution. Indeed, an averaging of distributions with different mean energies is well-known to overpopulate the tail of the distribution  \cite{beck1,lambi2}.
In the present case, the existence of these very energetic particles (as compared to the average) arises due to the fluctuations in the number of collisions. We have checked this effect by DSMC. To do so, we have focused on particles belonging to the tail of the distribution, i.e. the $N_t$ particles whose energy is larger than $k$ times the average, say $k=5$. Among these $N_t$ particles, we consider the proportion $p_t$ of particles  having performed less collisions than the average $t$, $p_t=N^{-}_t/N_t$. The time evolution of $p_t$ shows a power-law, i.e. slow, relaxation that confirms the important role played by the observed fluctuations in the high energy tail  for very long times.

\section{Conclusion}

In this paper, we focus on the phase space properties arising in inelastic gases. To do so, we  study  a two-dimensional Inelastic Maxwell Model (IMM), which is a mean field approximation of the Boltzmann equation for inelastic hard discs. 
By performing DSMC simulations of the model, we show 
a statistical relation between number of {\em inelastic } collisions undergone by a particle and its average energy.
This relation leads to
complex non-equilibrium features, namely the broadening of the 
available volume in phase space, that is followed by a 
power-law shrinking of this volume. 
 This mechanism is specific to inelastic gases, where energy is dissipated at each collision, and has no counterpart in elastic gases. 
 Finally, let us stress that  this work opens perspectives for a phase space description of inelastic gases, e.g. the use of Liouville equations for inelastic hard spheres \cite{leon,dufty}.

{\bf Acknowledgements}
This work 
has been supported by European Commission Project 
CREEN FP6-2003-NEST-Path-012864 and COST P10 (Physics of Risk).

\end{document}